\documentclass[
&aps,
prd,
twocolumn,
&onecolumn
]{revtex4-2}

\newcommand{\beqa}{\begin {eqnarray}}
\newcommand{\eeqa}{\end {eqnarray}}

\usepackage{float}
\usepackage{natbib}
\usepackage{amsbsy}
\usepackage{amssymb}
\usepackage{amsmath}
\usepackage{graphicx}
\usepackage{mathrsfs}
\usepackage[caption = false]{subfig}
\usepackage{array}
\usepackage{mathrsfs}% for script P
\usepackage{xfrac}
\usepackage{booktabs}
\usepackage{amsthm}
\usepackage{hyperref}
\hypersetup{colorlinks=true, citecolor=blue, urlcolor=blue, linkcolor=blue}

\usepackage{mathtools} % Need for cases.
\usepackage{graphicx}% Include figure files
\usepackage{dcolumn}% Align table columns on decimal point
\usepackage{bm}% bold math
\usepackage{etoolbox}
\apptocmd{\sloppy}{\hbadness 10000\relax}{}{}

\usepackage{xcolor}
\usepackage{amsthm}
\usepackage{pifont}

\usepackage{tikz}
\usetikzlibrary{positioning}
\tikzstyle{box} = [rectangle, draw=black, rounded corners, minimum width=3cm, minimum height=1cm, align=center]
\tikzstyle{arrow} = [thick, -{Latex[length=3mm, width=2mm]}]

\usepackage [english]{babel}
\usepackage [autostyle, english = american]{csquotes}
\usepackage{natbib}
\MakeOuterQuote{"}
\usepackage{enumerate}

\usepackage{ifpdf}
\usepackage[T1]{fontenc}
\usepackage[english]{babel}
\usepackage[autostyle, english=american]{csquotes}

\usepackage{amsmath,amssymb,amsbsy,mathtools,amsthm, bm}
\usepackage{mathrsfs}
\usepackage{graphicx}
\usepackage{dcolumn}
\usepackage{tabularx,booktabs, array}
\usepackage{natbib}
\usepackage{xcolor}
\usepackage{tikz}
\usetikzlibrary{shapes.geometric, arrows.meta, positioning}

\usepackage{subcaption}
\usepackage{caption}

\usepackage{etoolbox}
\apptocmd{\sloppy}{\hbadness 10000\relax}{}{}

\usepackage{hyperref}
\hypersetup{colorlinks=true, citecolor=blue, urlcolor=blue, linkcolor=blue}

\tikzstyle{box} = [rectangle, draw=black, rounded corners, minimum width=3cm, minimum height=1cm, align=center]
\tikzstyle{arrow} = [thick, -{Latex[length=3mm, width=2mm]}]

\def\be{\begin{equation}}
\def\ee{\end{equation}}

\def\be{\begin{equation}}
\def\ee{\end{equation}}

\begin{document}
\title{The dynamics of background  cosmological evolution and structure formation in phase space:  a semi-cosmographic reconstruction}% Force line breaks with \\
\author{Pankaj Chavan}\email[E-mail: ]{chavanpankaj09@gmail.com}
\author{Tapomoy Guha Sarkar}
\email[E-mail: ]{tapomoy1@gmail.com}
\affiliation{Department of Physics, Birla Institute of Technology and Science - Pilani, Rajasthan, India}
\author{Anjan A Sen}
\email[E-mail: ]{aasen@jmi.ac.in}
\affiliation{Centre for Theoretical Physics, Jamia Millia Islamia, New Delhi-110025, India}

\begin{abstract}
The Baryon Acoustic Oscillation (BAO) feature, imprinted in the transverse and radial clustering of dark matter tracers, enables the simultaneous measurement of the angular diameter distance $D_A(z)$ and the Hubble parameter $H(z)$ at a given redshift. Further, measuring the redshift space anisotropy (RSD) allows us to measure the combination $f_8(z)\equiv f\sigma_8(z)$. Motivated by this,  we simultaneously study the dynamics of background evolution and structure formation in an abstract phase space of dynamical quantities $x=H_0 D_A/c$, $p = dx/dz$, and $f_8$. We adopt a semi-cosmographic approach, whereby we do not pre-assume any specific dark energy model to integrate the dynamical system. The  Luminosity distance is expanded as a Pad\'e rational approximation in the variable $(1+z)^{1/2}$. The dynamical system is solved by using a semi-cosmographic equation of state, which incorporates the dark matter density parameter along with the parameters of the Pad\'e expansion. The semi-cosmographic $D_A(z), H(z)$ and  $f\sigma_8(z)$, thus obtained, are fitted with BAO and RSD  data from the  SDSS IV. The reconstructed phase trajectories in the $3D$ $(x,p,f_8)$ space are used to reconstruct some diagnostics of background cosmology and structure formation. At low redshifts, a discernible departure from the $\Lambda$CDM model is observed. The geometry of the phase trajectories in the projected spaces allows us to identify three key redshifts where future observations may be directed for a better understanding of cosmic tensions and anomalies.

\end{abstract}

\maketitle

% \begin{keyword}
% Baryon Acoustic Oscillation \sep Dark energy \sep Cosmography \sep Structure formation
% \end{keyword}

% \end{frontmatter}

\section{Introduction} 
The $\Lambda$CDM concordance model \cite{BULL_et_al_2016_Beyond_LCDM} has been widely successful in explaining diverse cosmological observations such as - the anisotropies in the cosmic microwave background (CMB) radiation \cite{Page_2003}, baryon acoustic oscillations (BAO) \cite{Eisenstein_1998_Baryonic_features, Eisenstein_2005, Matsubara_2004_BAO, Meiksin_1999_BAO}, and the large-scale structure of the Universe \cite{BERNARDEAU_2002, BULL_et_al_2016_Beyond_LCDM}, etc. By considering the cosmological constant $\Lambda$ as the simplest dark energy candidate \cite{Ratra-Peebles_1988, Padmanabhan_2003, amendola_tsujikawa_2010, bamba2012dark}, it offers the simplest explanation for the accelerated expansion of the universe \cite{Perlmutter_1997, Spergel_2003, Hinshaw_2003, scranton2003physical, Eisenstein_2005, McDonald_2007, Riess_2016}. However, despite its remarkable success, several theoretical challenges and observational tensions plague this cosmological framework \cite{BULL_et_al_2016_Beyond_LCDM, perivolaropoulos_2022_lcdm_challenges}. Combining DESI results with other data sets also seems to point towards dynamical dark energy \cite{DESI_Colab_2025_DR2}. 
While the standard interpretation of the cosmological constant as \textit{vacuum energy} has always faced a theoretical problem of extreme fine-tuning  \cite{weinberg1989cosmological, copeland2006dynamics, J_Martin_cosmo_cnst, Sola_2013_cosmo_cnst, perivolaropoulos_2022_lcdm_challenges}, the recent issues with the $\Lambda$CDM model mostly originate from observations. 

The Hubble-tension  points towards a consistent $(> 4\sigma)$ discrepancy between the value of $H_0$ measured using the local distance ladder with Cepheids (SH0ES) \cite{riess2021cosmic} and the implicit high redshift measurements from the CMBR \cite{PLANCK18_COSMO_params_Aghanim_2020}, Baryon Acoustic Oscillation (BAO) observations \cite{Eisenstein_2005, anderson2014clustering, beutler20116df, Alam_2021_SDSS, DESI_Colab_2025_DR2}, Big Bang nucleosynthesis \cite{fields2011primordial} and supernova (SNIa) observations \cite{sandage2006hubble, riess2021cosmic, beaton2016carnegie, freedman2020calibration,blakeslee2021hubble}. 
Also, the value of the growth parameter combination $S_8 \equiv \sigma_8 ( \Omega_{m0}/ 0.3 )^{0.5}$  as inferred from high redshift CMB (for example Planck TT, TE, EE+lowE \cite{aghanim2020planck}) observations appears to be in $2-3\sigma$ tension \cite{Karim_2025_sigma8} with local measurements using low redshift probes like weak gravitational lensing \cite{abbott2022_DESY_weaklensing, Porredon_DESY_lensing_2022, Pandey_DESY_gal_gal_lensing_2022, Hikage_2019_CosmicShear_PS, Karim_2025_sigma8} and galaxy clustering \cite{Alam_2021_SDSS, anderson2014clustering}. Further, attempts to resolve the Hubble tension by modifying the dark-energy/dark-matter sector have often aggravated the $S_8$-tension.

A wide diversity of dark energy models aims to tackle the persistent theoretical issues surrounding the nature of cosmic acceleration at late times.
These frameworks encompass a variety of dynamical scalar field theories \cite{Ratra-Peebles_1988, Steinhardt_1998, PhysRevLett.82.896, scherrer2008thawing},  extend to scenarios involving modifications of General relativity \cite{Khoury_2004, Hu_2007, nojiri2007introduction, Starobinsky_2007, amendola_tsujikawa_2010}  or introduce couplings within the dark sector \cite{amendola2000coupled, Pourtsidou_2013_Coupled_DE_DM}. Despite these attempts, the theoretical and observational landscapes still remain fraught, and despite the accompanying discomfort, the $\Lambda$CDM model is still the widely accepted cosmological paradigm. 

The absence of a compelling alternative has increasingly motivated the development of data-driven, model-independent approaches. Model-agnostic approaches include machine learning techniques like Gaussian process reconstruction \cite{Holsclaw_2011_GPR,Shafieloo_2012_GPR, Jesus_2024_GPR, Dinda2024_GP_cosmography,Velazquez_mukherjee_GPR_2024, Purba_Mukherjee-Anjan_sen_GPR_2024, Dinda_2025}. These data-driven methods do not rely on any dark energy model, making them useful for potentially resolving model degeneracies. However, extreme reliance on data makes it impossible to incorporate any known aspects of cosmological evolution (like our understanding of radiation and the cold dark matter sectors). 
 
Cosmography \cite{Weinberg_1972_cosmography} offers another model-independent method, focusing on describing the Universe through its kinematic properties, rather than assuming any specific dynamical model. In traditional cosmography, the physically measurable cosmological quantities like the  Hubble expansion rate or cosmological distances are expressed as a Taylor series expansion,  where the expansion coefficients are related to kinematic quantities \cite{Visser_2015_cosmography, Dunsby_Luongo_2016_cosmography, Capozziello_2019_cosmography, Busti_2015_cosmography, Visser2005-cosmography, Yang_Aritra_banerjee_2020_cosmography, Aviles_2013_cosmography, Aviles_bravetti_cosmography_2013,Aviles_2012_cosmography_y_variable, Cattoen_2007_convergence, Capozziello_2020_cosmography, P_Chavan_2025_semiCosmography} at the present epoch. These quantities include the Hubble constant $H_0$, deceleration $q_0$, jerk $j_0$, snap $s_0$ and lerk $l_0$, etc. 
which are constrained using the observed cosmological data. Taylor series expansions suffer from convergence problems,  thereby forcing the method to be applicable only at small redshifts $(z<1)$ \cite{Capozziello_2019_cosmography, Cattoen_2007_convergence, Capozziello_2020_cosmography, Gruber_Luongo_2014_cosmography, Lobo_2020, P_Chavan_2025_semiCosmography}.  This makes the Taylor expansion a bad choice for cosmography,  since most of the current cosmological observations are at redshifts larger than the radius of convergence.  The  $y$-variable method \cite{Aviles_2012_cosmography_y_variable, Cattoen_2007_convergence, Capozziello_Ruchika_Anjan_2019_cosmography} or Pad\'e approximation \cite{Pourojaghi2022, Petreca_2024, Wei_2014_cosmography, Gruber_Luongo_2014_cosmography, Capozziello_Ruchika_Anjan_2019_cosmography, Aviles_2014_cosmography, Mehrabi_2018_cosmography, Rezaei_2017_cosmography, Zhou_2016_cosmography, Liu_2021_cosmography, dutta2020beyond, Capozziello_2019_cosmography, Capozziello_2018_cosmography, Benetti_2019_cosmography, Pade_1892} are used instead of simple Taylor series expansion for enhanced radius of convergence. However, recent studies have shown that though the $y$-variable method solves the convergence issue, it provides a poor fit to the high redshift data as compared to cosmography with Pad\'e approximation \cite{Capozziello_Ruchika_Anjan_2019_cosmography},  making the latter approach more reliable. 

Pad\'e cosmography adopts a ratio of two polynomials in redshift to express the evolution of observable quantities.  In this approach,  a dark energy equation of state $w(z)$ may be reconstructed,  but only for known values of the density parameters $\Omega_i$ for the other non-dark energy components of the Universe. The values of the density parameters $\Omega_i$ are either constraints from other cosmological probes,  or they are obtained by relating the constrained cosmographic parameters of the Pad\'e expansion (like $q_0, j_0, s_0, l_0$, etc.)  with the parameters of the $\Lambda$CDM model. Such an attempt to associate the kinematic parameters with the parameters of the $\Lambda$CDM model can become challenging for non-trivial Pad\'e approximations (like Pad\'e approximation in ${\rm log}(1+z)$ \cite{Lusso_2019} or $\sqrt{1+z}$ \cite{Saini_2000}). Further, whatever the case may be, the choice of $\Omega_i$ should be such that the dark energy equation of state does not diverge for the given set of cosmographic parameters. In such cases, the choice of $\Omega_i$ can introduce bias in the reconstructed $w(z)$ and may not reveal the true nature of dark energy. This problem can be avoided by using a semi-cosmographic approach \cite{P_Chavan_2025_semiCosmography}. In this method,  an intermediate semi-cosmographic equation of state parameter $w(z)$ is introduced, which depends on the Pad\'e expansion parameters along with $\Omega_i$. This semi-cosmographic equation of state is subsequently used to compute observable quantities, which are then fitted with data to put constraints on  $\Omega_i$ along with the parameters of the Pad\'e expansion. 

The Hubble expansion rate $H(z)$, angular diameter distance $D_A(z)$ are crucial observables of background cosmology. These quantities are measured using cosmic chronometers \cite{Jimenez_2002_CC, Stern_2010_CC, Zhang_2014_SDSS_CC},  BAO \cite{Eisenstein_2005, anderson2014clustering, beutler20116df, Alam_2021_SDSS, DESI_Colab_2025_DR2} or Supernova observations \cite{sandage2006hubble, riess2021cosmic, beaton2016carnegie, freedman2020calibration,blakeslee2021hubble}. The quantity  $f\sigma_8(z)$, which quantifies the growth of structures, is measured from the clustering of dark matter tracers, where it appears due to redshift space distortion. 
This motivates us to study the dynamics of background evolution and structure formation in an abstract 3-dimensional phase space of these quantities. 
In this paper, we conjointly constrain $\Omega_{m0}$, $\sigma_8$, $H_0$ and other Pad\'e parameters by adopting the semi-cosmographic framework \cite{P_Chavan_2025_semiCosmography}. We use data on the measured covariance of these three quantities to reconstruct the cosmological evolution in phase space.  The reconstructed evolution in the 3-D phase space is
then compared with theoretical phase trajectories for a few known models. Further, the phase picture allows us to find three key redshifts where future observations should be directed.

\section{Formalism}

In the standard paradigm of cosmology, 
important observables fall under two categories - the quantities related purely to the background evolution of the homogeneous and isotropic Universe and the quantifiers of inhomogeneities and structure formation.
 
The observable dynamical quantities pertaining to the background evolution are the Hubble expansion rate and cosmological distances.  Diverse observational cosmological probes aim to measure the Hubble expansion rate (using cosmic chronometers \cite{Jimenez_2002_CC, Stern_2010_CC, Ratsimbazafy_2017_diff_age, Zhang_2014_SDSS_CC}) and cosmological distances (using standard candles like SNIa \cite{sandage2006hubble, riess2021cosmic, beaton2016carnegie, freedman2020calibration,blakeslee2021hubble}) directly. 
Measuring the Baryon Acoustic Oscillations (BAO) feature in the transverse and radial clustering of dark matter tracers allows the simultaneous measurement of the Hubble expansion rate $H(z)$ and the angular diameter distance $D_A(z)$ at a  certain redshift $z$, using the 
sound horizon $r_d$ at the drag epoch as a  standard ruler \cite{Eisenstein_1998_Baryonic_features, Eisenstein_2005}.
While expansion rate and distances can be measured independently, they are related to each other in an FLRW cosmology. For example,   $H(z)$ and  $D_A(z)$ are related through the relation
\begin{equation}
    D_A(z) = \frac{c}{H_0}\frac{1}{\sqrt{|\Omega_{k0}|}}\frac{1}{1+z}S_k\left(H_0\sqrt{|\Omega_{k0}|}\int_{0}^{z}\frac{dz'}{H(z')}\right)
\end{equation}
where,
\begin{equation}
S_k(y) = 
\begin{cases}
\sin\left(y\right) & \text{if } \Omega_{k0} < 0 \ \\
y & \text{if } \Omega_{k0} = 0 \ \\
\sinh\left(y\right) & \text{if } \Omega_{k0} > 0 \
\end{cases}.
\end{equation}
Noting that the Hubble expansion rate $H(z)$ is essentially related to the derivative of $D_A(z)$, it is convenient to define  dimensionless variables $x(z)$ and $p(z)$ as
\be
    x (z) = \frac{H_0}{c}D_A(z)~~~\text{and}~~~p(z) =  \frac{d x}{dz}.
    \label{eq:x(z) and p(z)}
\ee

The general relation between these variables in the framework of general relativity with the FLRW metric is thus given by a consistency condition 
\begin{equation}
    x+p(1+z) =\frac{K(z)}{E(z)}
    \label{eq:consistency}
\end{equation}
where the function $K(z)$, which accounts for any constant but a non-zero spatial curvature,  is given by 
\begin{equation}
K(z) = 
\begin{cases}
\sqrt{1 -|\Omega_{k0}| x^2(1+z)^2} & \text{if } \Omega_{k0} < 0  \\
1 & \text{if } \Omega_{k0} = 0  \\
\sqrt{1 +|\Omega_{k0}| x^2(1+z)^2} & \text{if } \Omega_{k0} > 0 
\end{cases}
\end{equation}
and $E(z) = H(z)/H_0$. 
The dynamics of the expanding Universe  described by the second-order dynamical Friedmann equation can be equivalently written as a pair of first-order equations in the dynamical variables $(x,p)$ as 
\begin{eqnarray}
x'&=& p \\
p' &=& -\frac{1}{1+z} \left[ 2p + \frac{{E}'}{ E^2}K - \frac{K'}{E}  \right].
\end{eqnarray}
Here, and throughout the paper, the \textit{prime} shall denote derivative with respect to the redshift $z$. 
The entire background evolution can now be seen as the phase trajectory $p(x)$.
To integrate the above system of equations, we need to know the functions $E(z)$, $K(z)$, $E'(z)$, and $K'(z)$.
For a multi-component  Universe with curvature, dark matter, and dark energy, we may write
\be E(z) = \sqrt{\Omega_{k0} (1+z)^2 + \Omega_{m0} (1+z)^3 +\Omega_{\phi 0} f_{\phi}(z)}
\label{eq:E(z)^2} \ee
where $f_{\phi}(z)$ imprints the dynamics of dark energy. Observations indicate that the Universe is spatially flat, whereby we assume that $\Omega_{k0} =0$.
If dark energy is treated as a fluid with an equation of state parameter $P/ \rho = w(z)$, we have 
\be
    f_{\phi}(z) = exp\left(3 \int_{0}^{z}\frac{1+w(z')}{1+z'}dz'\right).
\ee
This gives us an autonomous system of non-linear coupled differential equations
\begin{eqnarray}
 x'&=& p  \nonumber \\
{p}'&=&-\frac{2p}{1+z} - \frac{3}{2} \frac{ \left( 1 + w \right) (x + p + pz)}{(1+z)^2} \nonumber  \\  && \hspace{0.425 cm} + \frac{3}{2} \Omega_{m0} w(1 + z)(x + p + pz)^3  \nonumber \\
z'&=& 1
\label{eq:autonomous_bgrnd_evln}
\end{eqnarray}
which can be integrated for a given dark energy equation of state $w(z)$  using the initial conditions $(x_0, p_0, z_0) = ( 0, 1, 0)$.

The dynamics of cosmological structure formation is encapsulated in the temporal part of the overdensity field $\delta$. The linearized theory of gravitational instability in an expanding Universe gives a growing solution for the density fluctuations. The second-order differential equation governing this growing mode $D_+(z)$ of density perturbations is given by
\be
    D''_+ + \left[\frac{E'(z)}{E(z)} - \frac{1}{1+z}\right]D'_+ = \frac{3}{2}\frac{\Omega_{m0}(1+z)}{E(z)^2}D_+.
    \label{Eq:Growth_Rate_Diff_Eqn}
\ee
where we have assumed that the dark energy does not cluster, and dark matter is the key driver of structure formation. On large scales, baryonic matter is expected to trace the overdensity field of dark matter. We note that the solution to this differential equation depends on the background evolution through $E(z)$ and $E'(z)$.

The quantity pertaining to this growing mode is the logarithmic growth rate 
\be f(z) = -(1+z)\frac {d \ln D_+}{dz}. \ee 
This quantity appears in the measured clustering of dark matter tracers (the two-point function or the power spectrum) in redshift space, and arises due to the non-Hubble contribution to redshift coming from peculiar flow.
In actual observation of redshift space distortions (RSD) in tracer clustering (like Kaiser effect in galaxy surveys) \cite{Beutler_2012_6dF_growthRate} the combination $f(z)\sigma_8(z) \equiv f_8(z)$ is measured \cite{Alam_2017_SDSS_III, Alam_2021_SDSS} where $\sigma_8(z) = \sigma_8 D_+(z)/ D_{+}(0)$ is the root mean square of the amplitude of density perturbations averaged over the spheres of radius $8h^{-1}~\text{Mpc}$.
The quantity  $f_8(z)$ is sensitive to both the growth of structure and the amplitude of fluctuations, making it a powerful cosmological probe. 

The second-order differential equation Eq. \ref{Eq:Growth_Rate_Diff_Eqn} can be cast as an autonomous system of equations
\begin{eqnarray}
    \sigma_8'(z)&=&-\frac{f_{8}(z)}{1+z}\nonumber \\
    f_{8}'(z)&=&\frac{3}{2}\frac{\Omega_{m0}(1+z)^2}{E^2(z)}\left(w(z)f_{8}(z)-\sigma_{8}(z)\right) \nonumber  \\  && \hspace{0.42 cm} + \frac{f_{8}(z)}{2}\left(\frac{1-3w(z)}{1+z}\right) \nonumber \\
    z'&=&1
    \label{eq:autonomous_structure_frmn}
\end{eqnarray}
and can be solved to obtain $f_8(z)\equiv f(z)\sigma_8(z)$ with initial conditions 
\begin{equation}
    \left(\sigma_{8}(z_{init}), f_{8}(z_{init}), z_{init} \right) =\left(\frac{\sigma_8}{1+z_{init}}, \frac{\sigma_8}{1+z_{init}}, z_{init}\right)
\end{equation}
where $z_{init}$ is some early redshift corresponding to the matter-dominated epoch.

\begin{figure*}[ht]
\centering
\includegraphics[height=11cm, width=15.7143cm]
{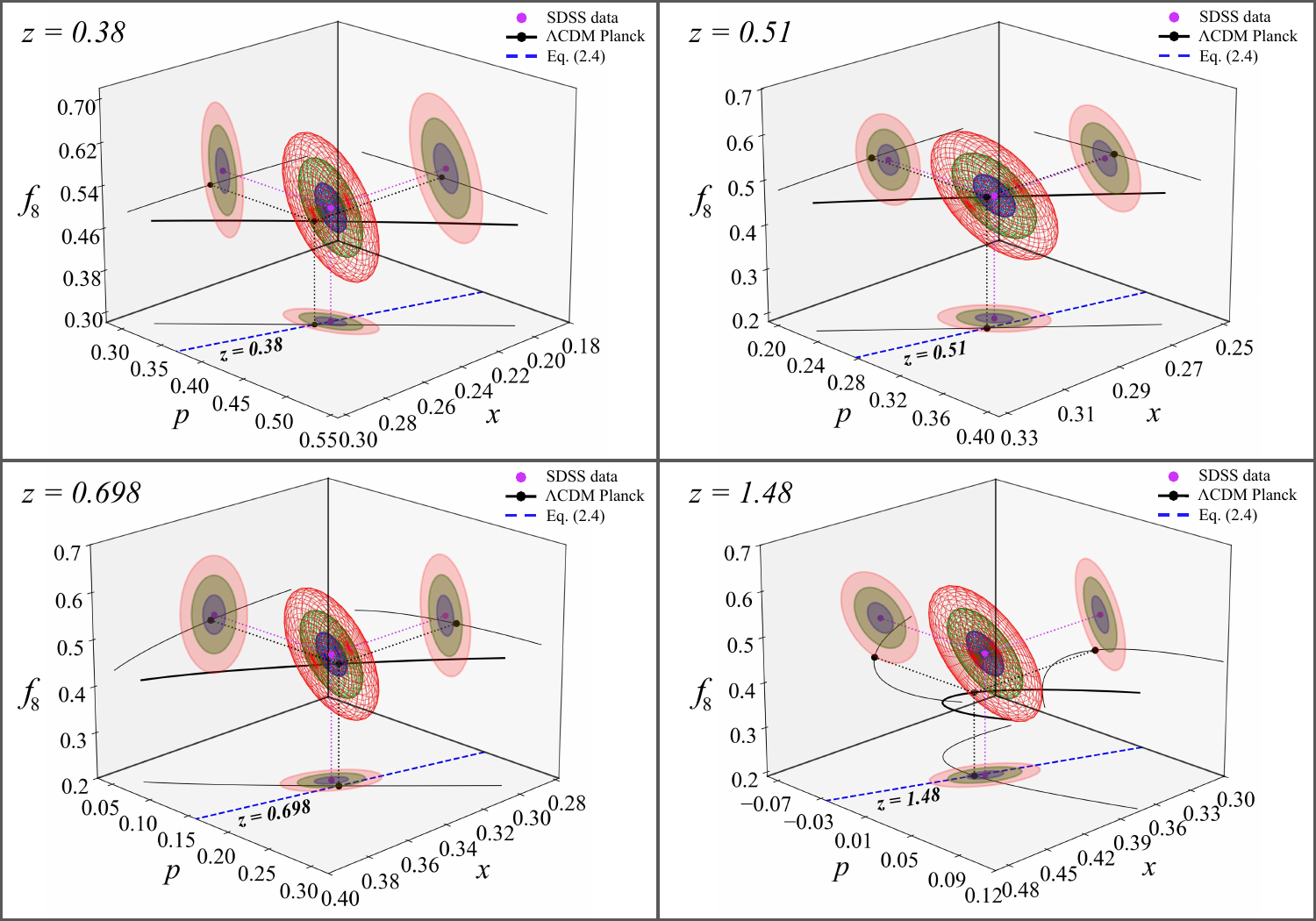}
\captionsetup{justification=justified, singlelinecheck=false}
\caption{The cosmological evolution in 3-D $(x,p,f_8)$ space for $\Lambda$CDM model with Planck18 parameters is shown along with the projections on the $(x, p)$, $(x, f_8)$ and $(p, f_8)$ spaces. The 3D ellipsoids shown above represent the SDSS IV covariance data measured at their respective redshifts. We also show the straight line of consistency in Eq.  \ref{eq:consistency} in the $(x,p)$ plane.}
\label{fig:3D_phase_space_LCDM_P18_with_SDSS_Covariances}
\end{figure*}

The solution of the autonomous system in Eq. \ref{eq:autonomous_bgrnd_evln} and Eq. \ref{eq:autonomous_structure_frmn} gives us the evolution of the dynamical quantities ($x(z)$, $p(z)$, $\sigma_{8}(z)$, $f_8(z)$) in an abstract 4-D phase space. However, the directly measurable dynamical quantities, namely the Hubble expansion rate, the Angular diameter distance, and the growth rate of density fluctuations, 
are related to the variables $(x, p, f_8)$.
The space spanned by these three dynamical variables forms a 3-D projection of the full 4-D phase space. 
We are interested in the dynamics in this 3-D space due to its direct relevance to observations. 

It is a standard practice to study these three observables as a function of $z$ separately. Ruling out cosmological models based on data, then requires us to study three different evolutions $x(z)$, $p(z)$, and $f_8(z)$ separately. However, if we are initially unaware of the sensitivity of the three dynamical quantities $(x, p, f_8)$ to different dark energy models, it would be naturally more convenient to study all three of them simultaneously in the abstract phase space.

In this work, we have considered the 3-D phase space of the dynamical variables $(x, p, f_8)$. Eliminating $z$ between these quantities gives a curve in the 3-D space that captures the cosmological evolution of all these measurable quantities simultaneously over the entire evolution history.
In this abstract phase space, any cosmological model corresponds to a unique curve. Every point on the curves refers to the values of $(x,p,f_8)$ for that model at some redshift $0 \leq z \leq \infty$.
Further, in this 3-D phase space approach,  the information about background evolution and perturbations is studied simultaneously. Many models that are indistinguishable from each other as regards their background cosmological evolution can be immediately distinguished by their sensitivity to structure formation. Moreover,  measured covariances between RSD data and BAO data make it imperative to see $(x,p,f_8)$ together simultaneously. Figure \ref{fig:3D_phase_space_LCDM_P18_with_SDSS_Covariances} shows the BAO+RSD data in the 3D $(x,p,f_8)$ space at the four observing redshifts $0.38, 0.51, 0.698$ and $1.48$. Assuming the data is Gaussian distributed, we present the covariance information as error ellipsoids around each measured data point in this 3D space. The projections on the $(x,p)$,  $(x,f_8)$, and $(p,f_8)$ planes are also shown.
We also plot the Planck $\Lambda$CDM prediction for $(x,p,f_8)$ at these observed redshifts.
If an ellipsoid has an equation of the form $F(x, p, f_8) = 0$, then one can algebraically ascertain if a model-predicted point ($x_{model}$, $p_{model}$, $f_{8model}$) is inside or outside the ellipsoid by checking if $F(x_{model}, p_{model}, f_{8model})  <0 $ or $>0$ respectively.

Figure \ref{fig:3Dphasespace} shows the dynamics of cosmic evolution in the $(x, p, f_8)$ space for different cosmological models. Each of these models has a different description for dark energy. 
As a small sample of the great diversity of dark energy models, we have considered the following models in this work:

\begin{itemize}
\item { \textbf{The $\Lambda$CDM model:} This is the simplest cosmological model with a constant DE equation of state parameter ($w=-1$). For the $\Lambda$CDM Planck model, we have considered the Planck \cite{aghanim2020planck} estimates of the model parameters, and for the $\Lambda$CDM SDSS model, we have considered the parameters listed in the lower row of Table \ref{tab:MCMC-constraints} (these parameters are obtained by fitting the $\Lambda$CDM model with SDSS data).}
\item { \textbf{The CPL model:} The DE equation of state $w_{\phi}(z)$ for the  CPL model \cite{CHEVALLIER_2001} is parametrized using two parameters $(w_0, w_a)$ and is given by 
\be w_{\phi}(z) = w_0 + w_a \frac{z}{1+z}. \ee
According to this two-parameter phenomenological para\-metrization, the dark energy equation of state evolves from $w_{\phi}(z)=w_0 + w_a$ (early universe) to $w_{\phi}(z)=w_0$ (present universe). For this model, we use the parameters $H_0 = 68.31 \text{~Km/s/Mpc}, w_0 = -0.957, w_a = -0.29, \sigma_8 = 0.820$ and $S_8 = 0.829$ from \cite{aghanim2020planck}.}

\item { \textbf{The thawing quintessence (TQ) model:} Thawing models are a class of scalar field models that mimic the cosmological constant in the past $\left(w \sim -1\right)$ and deviate slightly (becoming less negative) as the universe expands.
We use the following DE equation of state
\be
~~~~~~~~1+w = \frac{(1+w_0)\left[ \sqrt{1+\Omega_e} - \Omega_e \tanh^{-1}\left(\frac{1}{\sqrt{1+\Omega_e}}\right) \right]^2}{\left[ \frac{1}{\sqrt{\Omega_{\phi0}}} - \left(\Omega_{\phi0}^{-1} - 1\right) \tanh^{-1}(\sqrt{\Omega_{\phi0}}) \right]^2}
\ee
where $\Omega_e =  \left(\Omega_{\phi0}^{-1} - 1\right)\left(1+z\right)^3 $. 
We have used the parameters $w_0 = -0.9, \Omega_{\phi0}=0.7$
from \cite{PhysRevD.77.083515} and $\sigma_8=0.811$ from \cite{aghanim2020planck}.}

\begin{figure*}[ht]
\centering
\includegraphics[height=8.5cm, width=18cm]{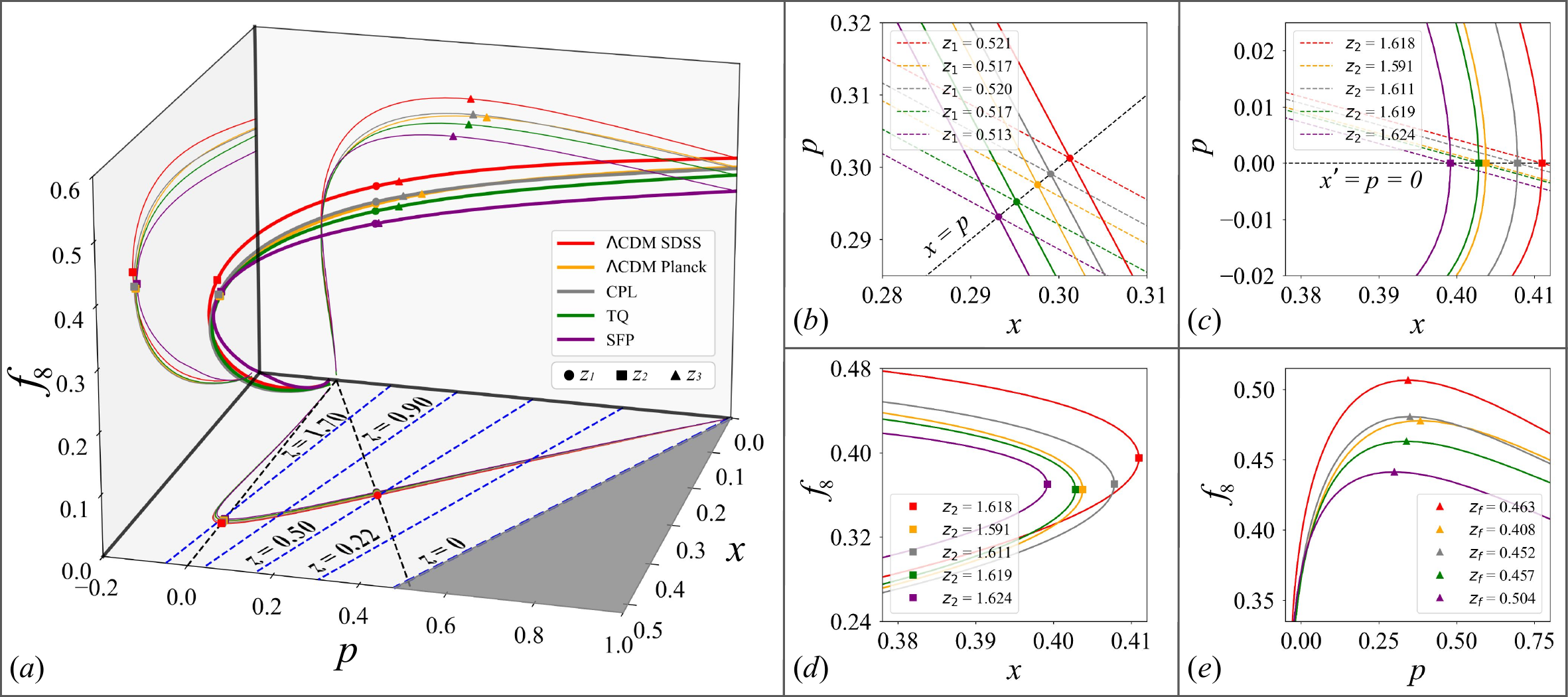}
\captionsetup{justification=justified, singlelinecheck=false}
\caption{ (\textit{a}) The cosmological evolution in 3-D $(x,p,f_8)$ space for different dark energy models is shown along with the projections on the $(x, p)$, $(x, f_8)$, and $(p, f_8)$ spaces. The blue dotted lines in the $(x,p)$ plane correspond to the consistency lines given in Eq. \ref{eq:consistency} for different redshifts. The intersection of the projected phase trajectories with the blue dotted lines gives the value of $(x,p)$ at the corresponding redshift. The diagonal black dotted line intersects the projected phase trajectories at points where $x=p$. (\textit{b}) This shows the part of the $(x, p)$ space where the $x=p$ line intersects the projected phase trajectories. For each model, this corresponds to a unique redshift $z_1$ at which $x(z_1) = p(z_1)$.  (\textit{c}) Shows the intersection of the projected phase trajectories in the $(x, p)$ plane with the line $p=0$. The intersection points correspond to the unique redshift $z_2$ for which $x'(z_2)= p(z_2) =0$. 
The dashed lines in (\textit{b}) and (\textit{c}) show the consistency lines Eq: \ref{eq:consistency} for each model, and are used to determine the redshifts.
(\textit{d}) Shows the $(x, f_8)$ plane with the marked points corresponding to $z_2$ for each model where $p$ vanishes. (\textit{e}) Shows the part of the $(p, f_8)$ space with the marked points on the projected phase trajectories at redshifts $z_f$ where $f_8$ is a maximum.}
\label{fig:3Dphasespace}
\end{figure*}

\item{ \textbf{Scale factor parametrization (SFP) model:} This model considers the scale factor parametrized with parameters $a_1$ and $B$ such that
\be a(t) = a_{1}^{2/B}[\sinh(t/\tau)]^{B}
\ee
where $\tau$ has dimensions of time.
We adopt a DE equation of state of the form 
\be
w(z) = \frac{w_T(z)E^2(z)}{E^2(z)-\Omega_{m0}(1+z)^3}
\ee
where
\be 
w_T(z)=\frac{2A}{3B}\frac{(1+z)^{2/B}}{A(1+z)^{2/B}+(1-A)} - 1
\ee 
with
\be
A = \frac{a_{1}^{2/B}}{\left(1 + a_{1}^{2/B}\right)}
\ee
and
\be
E^2(z) = \left[A(1+z)^{2/B} + (1-A)\right]. 
\ee
We adopt the parameters $A=0.39$, $B = 0.72$, $\Omega_{m0}=0.25$ and $\sigma_8=0.85$ from \cite{scalefactorparam}.}

\end{itemize}

Redshift is a parameter in the 3D curves showing the cosmological evolution for different models. However, we must crucially note that each curve corresponding to a certain model is a different Universe altogether, and 
comparison of Cosmological evolution in the 3D space between different models has to be done for a given redshift. Thus, the point of intersection of two curves corresponding to two models does not indicate that these two models agree, as the point of intersection in general corresponds to different redshifts for the two models. Any comparison between models or any comparison between data and a model should be done at given fixed redshifts and thereby identifying points in the 3D space for those redshifts. The visual proximity of curves in 3D for different models does not necessarily mean that these models are similar. The projection of the trajectories onto the 2D $(x,p)$ plane, however, allows us to compare models as the intersection of the projected curves in this 2D plane with the consistency straight lines in Eq. \ref{eq:consistency} allows us to compare models at the given redshift.

In Figure \ref{fig:3Dphasespace}(\textit{a}), each of the curves in the $3$ dimensional ($x$,$p$,$f_8$) space corresponds to the complete evolution for each of the models from $z=\infty$ to $z=0$.  The projections on the $(x,p)$, $(x,f_8)$ and $(p, f_8)$ planes are also shown. 
While each of the projections has partial cosmological information, the $(x,p)$ projection is particularly important since the dynamical pair $(x,p)$ is measured in BAO observations. The shaded gray region in the $(x,p)$ plane corresponds to $z<0$. Since $D_A$  increases with redshift and then decreases after reaching a peak value, it implies that the accessible region of the $(x,p)$ phase space is bounded. 
For every cosmological model,  the trajectory projected on the  $(x,p)$ plane starts at $(0,1)$ corresponding to $z=0$ and ends up at $(0,0)$ for $z=\infty$.
The dotted blue lines in the $(x,p)$ plane represent the consistency condition between $D_A(z)$ and $H(z)$ given in Eq. \ref{eq:consistency}, for a given redshift. 
The intersection of the curves in 3-D  with the vertical planes standing on the blue dotted lines, thus gives the actual $(x,p,f_8)$ for the corresponding redshift.
There are two key redshifts $(z_1, z_2)$. These correspond to the points where the lines $x=p$ and $p=0$ intersect with the projected phase trajectories in the $(x,p)$ plane, respectively. At $z_1$ we have  $x(z_1) = p(z_1) $, and one has full information of background cosmology from a measurement of either $x(z_1)$ or $p(z_1)$. Since there is a perfect correlation between $x$ and $p$,  the error ellipses on the projected plane at this redshift become a line. These points on the projected phase trajectories at redshifts $z_1$ are shown in the Figure \ref{fig:3Dphasespace}(\textit{b}).
For the chosen fiducial parameters for the models
$\Lambda$CDM (SDSS), $\Lambda$CDM (Planck), CPL, TQ, and SFP, we find that $z_1 = 0.521, 0.517, 0.520, 0.517$ and $0.513$ respectively.

If we leave the singular point at the big bang corresponding to $z=\infty$, where $(x,p) = (0,0)$,  the second key redshift $z_2$,  at which $p(z_2)=0$, corresponds to the maxima of $D_A(z)$. Again, at this important redshift, only the measurement of $x(z_2)$ suffices, and $x(z_2)$ is directly related to $E(z_2)$. The phase point for different models at redshift $z_2$ are shown in the projected $(x,p)$ and $(x,f_8)$ planes in Figure \ref{fig:3Dphasespace}(\textit{c})(\textit{d}). We find that $z_2= 1.618, 1.591, 1.611, 1.619, 1.624$ for the models $\Lambda$CDM (SDSS), $\Lambda$CDM (Planck), CPL, TQ, and SFP, respectively.

There is also a redshift $z_3$ at which $p$ attains a minimum value. At this redshift $p'(z_3) =0$. This redshift is typically higher than the regime where dark energy dominates. Different dark energy models have the same behaviour at a high redshift in the matter-dominated epoch at around $z_3$. Hence, we do not focus our attention on $z_3$.

The two important redshifts $(z_1, z_2)$ shall always exist for every cosmological model. It is useful to find these redshifts at which observations may be focused for probing into the nature of dark energy.

If we look at the projections of the phase portrait on the $(x, f_8)$ and $(p, f_8)$ planes, we recognize that there is a crucial redshift $z_f$ for which $f_8(z)$ is a maximum.
Thus, ($x(z_f)$, $p(z_f)$, $f_8(z_f)$) is the highest point that the phase trajectory attains for a given model. It corresponds to 
$f_8'(z_f) =0$. From Eq. \ref{eq:autonomous_structure_frmn}, this condition is satisfied when 
\be 
\frac{\Omega_{m0} (1 + z_f)^3}{E^2(z_f)} = \frac{w(z_f) - 1/3}{w(z_f) - 1/f(z_f)}.
\ee
Interestingly, the left-hand side of this equation for $z_f$ depends only on background cosmology, while the right-hand side is sensitive to structure formation through $f(z_f)$. Thus, at this redshift, $f$ is determined by the background quantities, and the error of $f$ shall be related entirely to the errors in the background quantities $x$ and $p$.
In this work, we also consider this redshift $z_f$ at which different dark energy models may be studied.
Figure \ref{fig:3Dphasespace}(\textit{e}) shows the projected phase trajectories for the 
$\Lambda$CDM (SDSS), $\Lambda$CDM (Planck), CPL, TQ, and SFP models in the $(p, f_8)$ space, and the points on the projected phase trajectories corresponding to $z_f = 0.463, 0.408,  0.452,  0.457, 0.504$ respectively are shown.

For a given dark energy equation of state $w(z)$, the 3-D phase trajectory can be obtained. Constraining a dark energy model corresponds to reconstructing the 3-D phase trajectory. Given the great diversity of competing dark energy models, we adopt a more data-driven approach.
In this work, we do not {\it a priori} assume any specific dynamical model $w(z)$ and adopt a cosmographic route.

\subsection*{The semi-cosmographic reconstruction}
In a cosmographic approach, no assumption is made about the dynamics of dark energy, and no model $w(z)$ is assumed. Instead, observable quantities are expanded as a function of redshift. The expansion coefficients are then often related to kinematic quantities. 

In this work we begin with a Pad\'e approximation for the Luminosity distance $D_L(z)$, written in terms of variable $\xi = \sqrt{1+z}$ as \cite{Saini_2000}
\be
    D_L^{\cal P}(z) = \frac{2c}{H_0}\left [ \frac{  \xi^4 - \alpha  \xi^3 -( 1- \alpha )\xi^2 }{ \beta \xi^2 + \gamma  \xi + 2 - \alpha  - \beta - \gamma }   \right ].
\label{eq:pade_R}
\ee
This is not the usual form of writing a Pad\'e approximant. The choice of this function is motivated by the fact that it accommodates a wide range of models. In addition, $H^{\cal P}(z)$, obtained by differentiating $D_L^{\cal P}(z)/(1+z)$ yields the desired asymptotic behavior $H^{\cal P}(z) \rightarrow H_0$ in the 
limit $z \rightarrow 0$. 
It also reproduces the  form of $H^{\cal P}(z)$ for the matter dominated Universe  at high redshifts  with ${H^{\cal P}(z)}^2\rightarrow H_0^2 \left(\frac{\beta^2}{ \alpha \beta + \gamma}\right)^2 (1 + z)^{3}$ when $z>>1$ \cite{Saini_2000}.

The Pad\'e approximated $D_L^{\cal P}(z)$
gives us the corresponding Pad\'e approximated angular diameter distance and Hubble parameter as 
\be
    D_A^{\cal P}(z) = \frac{D_L^{\cal P}(z)}{(1+z)^2} ~~~\&~~~  H^{\cal P}(z)= c\left [\frac{d}{dz} \frac{D_L^{\cal P}(z)}{(1+z)}\right ]^{-1}.
\ee

In the standard cosmographic approach, $D_L^{\cal P}(z)$, $D_A^{\cal P}(z)$ and $H^{\cal P}(z)$ are fitted with observational data.  The constraints on the  Pad\'e expansion parameters $(\alpha, \beta, \gamma, H_0)$ are then used to reconstruct cosmological evolution of various dynamical quantities. 

We propose a semi-cosmographic formalism  by defining a semi-cosmographic dark energy equation of state $w(z)$ obtained from the Pad\'e approximated $H(z)$ as
\be
    w^{\cal P} (z) = \frac{
    \displaystyle \frac{2}{3}(1+z)\frac{d}{dz} \ln H^{\cal P}(z) - 1}{\displaystyle 1 - \left( \frac{H_{0}}{H^{\cal P}(z)} \right)^2 \Omega_{m0}(1+z)^3}.
    \label{eq:eos}
\ee
This semi-cosmographic equation of state parameter $w^P(z)$, depends on $\Omega_{m0}$ along with the  Pad\'e  parameters $(H_0, \alpha, \beta, \gamma)$. 
The dynamical systems given in Eq. \ref{eq:autonomous_bgrnd_evln} for background cosmological evolution and Eq. \ref{eq:autonomous_structure_frmn} for structure formation can be numerically integrated to obtain $(x(z), p(z), f_{8}(z))$ using $w^P(z)$. 
These quantities are then compared with data from BAO and RSD measurements to constrain the parameters $(H_0,\Omega_{m0},\sigma_8,\alpha,\beta,\gamma)$ and reconstruct some important dark energy diagnostics.

\begin{figure}[ht]
\centering
\begin{tikzpicture}[node distance=0.6cm]

\node (step1) [box] { $D_L^{\cal P} \left (z, \alpha, \beta, \gamma, H_0 \right )$ [Eq. \ref{eq:pade_R}] };
\node (step2) [box, below=of step1] { $D_A^{\cal P}(z) = \frac{D_L^{\cal P}(z)}{(1+z)^2} ~~~\&~~~  H^{\cal P}(z)= c\left [\frac{d}{dz} \frac{D_L^{\cal P}(z)}{(1+z)}\right ]^{-1} $};
\node (step3) [box, below=of step2] {$
w^{\cal P} (z) = \frac{
    \displaystyle \frac{2}{3}(1+z)\frac{d}{dz} \ln H^{\cal P}(z) - 1}{\displaystyle 1 - \left( \frac{H_{0}}{H^{\cal P}(z)} \right)^2 \Omega_{m0}(1+z)^3}
   $ [Eq. \ref{eq:eos}]};
\node (step4) [box, below=of step3] {$w^{\cal P} (z, H_0, \Omega_{m0}, \alpha, \beta, \gamma) \rightarrow [  x(z), p(z), \sigma_8(z), f_8(z)]$\\ by solving Eq. \ref{eq:autonomous_bgrnd_evln} and Eq. \ref{eq:autonomous_structure_frmn}};
\node (step5) [box, below=of step4] {Fit 
$[x(z), p(z), f_8(z)]$ with data \\ \&\\
constrain $(H_0, \Omega_{m0}, \sigma_8, \alpha, \beta, \gamma)$};

\draw [arrow] (step1) -- (step2);
\draw [arrow] (step2) -- (step3);
\draw [arrow] (step3) -- (step4);
\draw [arrow] (step4) -- (step5);

\end{tikzpicture}

\caption{The schematic flowchart for the semi-cosmographic method for estimating cosmological parameters.}
\label{fig:Flowchart}
\end{figure}

In this work, we have reconstructed the following diagnostics of background evolution and structure formation using the estimated parameters.

\begin{itemize}
\item { {\bf The dark energy EoS $w(z)$:} }
The DE equation of state parameter $w(z)$ is one of the most common quantifiers used to study the dynamics of DE. For the $\Lambda$CDM mode, $w(z)=-1$. Any departure from this value points towards a non-$\Lambda$CDM Universe with dynamical DE.

\item { \bf The ${\cal{O}}m$ diagnostic:} The ${\cal{O}}m$ parameter \cite{Sahni_2008_Om} defined as
\be
{\mathcal{O}}  m (z)= \frac{  E(z)^2  - 1}{\left(1+z\right)^3 -1}
\label{Eq: Om}
\ee
is another useful diagnostic used to quantify the nature of DE.  ${\mathcal{O}}m(z) = \Omega_{m0}$ for the $\Lambda$CDM  model. 
Further, for Phantom DE models,  ${\mathcal{O}}  m (z) < \Omega_{m0}$ and for Quintessence DE models ${\mathcal{O}}  m (z) > \Omega_{m0}$.
However, ${\mathcal{O}}  m (z)$,  suffers from  a divergence occurring at $z=0$.

\item{\bf DE evolution $f_{\phi}(z)$:} The function $f_{\phi}(z)$ that imprints the DE evolution is defined as
\begin{equation}
    f_{\phi}(z) = \frac{E(z)^2 - \Omega_{m0}\left(1+z\right)^3}{1 - \Omega_{m0}}
    \label{Eq: DE_evolution}.
\end{equation}
Since $f_{\phi}(z) =1$ for the $\Lambda$CDM model,  any deviation from unity points toward the dynamical nature of DE.

The above diagnostics are probes of background cosmology. We propose a diagnostic $\mathcal{S}(z)$ of dark energy from the evolving dynamics of structure formation.

\item {\textbf{Dynamics of growth rate of fluctuations $\mathcal{S}(z)$:}}

We define $\mathcal{S}(z)$ as
\be 
\mathcal{S}(z) = \frac{\Omega_{m0} (1 + z)^3}{E(z)^2} - \frac{w(z) - 1/3}{ w(z) - 1/f(z)}.
\label{Eq: s(z)}
\ee
This diagnostic has the following properties. 
\begin{enumerate}
\item At $z=z_f$ for which $f'_8(z_f) =0$, $\mathcal{S}(z_f) = 0$. Thus, the zero of the function $\mathcal{S}(z)$ corresponds to the maximum value of $f_8(z)$.
\item For the $\Lambda$CDM model, ${\mathcal S}(z)$ takes the form 
\be 
\mathcal{S}(z) \approx f(z)^r
- \frac{4}{3} \frac{f(z)}{1+f(z)}, \ee
with $r \approx 11/6$,  where,   $ f(z) $ takes the approximate form 
\be 
f \approx \left [ \frac{\Omega_{m0} (1 + z)^3}{\Omega_{m0} (1 + z)^3 + ( 1- \Omega_{m0}) } \right ]^{1/r}. \label{eq:lcdmf} \ee 
\item Irrespective of the cosmological parameters for $\Lambda$CDM model, $\mathcal{S}(z) \rightarrow 1/3$ for $z >>1$.
\item For any model dark energy $w(z)$ we have for $z>>1$, $\mathcal S(z) \rightarrow \frac{2}{3(1 - w(z))}$ 
\item 
As $z \rightarrow 0$, the parameter $\mathcal{S}(z)$  approaches the value $\left ( \Omega_{m0} - \frac{w(0) - 1/3}{w(0) - 1/f(0)} \right) $.
\end{enumerate}
\end{itemize}

The diagnostic $\mathcal{S}(z)$ involves a background quantity $E(z)$ and also a quantifier of structure formation $f(z)$. This makes it a unique dark energy probe.  At high redshifts $\mathcal S(z)$ is solely determined by $w(z)$; however, it is sensitive to $f(z)$ at low redshifts. 
The main importance of $S(z)$ is that the redshift $z_f$ where it vanishes, corresponds to the peak in $f_8(z)$.  Thus measurement of $S(z)$ around $z_f$  allows us to locate an important epoch in cosmic history where $f'_8(z) =0$. Identifying this redshift $z_f$ where $\mathcal{S}$ vanishes allows us to use BAO $(D_A, H)$ and RSD $f_8$ measurements to directly constrain $f(z_f)$ as is studied in \cite{P_chavan_2025_grthRt_Bias}.

In the next section, we discuss the observational probes that we shall subsequently use to 
(a) Constrain model parameters $(H_0, ~\Omega_{m0}, ~\sigma_8,  ~\alpha, ~\beta,~ \gamma)$ (b) reconstruct the cosmological evolution in the $(x,p,f_8)$ phase space (c) reconstruct dark energy diagnostics $(w(z),~\mathcal{O}m (z), ~f_{\phi}(z), ~\mathcal{S}(z))$ from background cosmology and structure formation respectively and (d) investigate the three important redshifts $(z_1, ~z_2, ~z_f)$ where focused attention is required.

\section{Observational Data}

\renewcommand{\arraystretch}{1.42}
\begin{table*}[ht]
\small
\centering
\setlength{\tabcolsep}{2pt}
\begin{tabular}{c c c c c c c c c c c }
\hline \hline
Tracer &$z_{eff}$ & $\widetilde{D}_{M}(z)$  & $\widetilde{D}_{H}(z)$ & $\widetilde{D}_{V}(z)$ & $f\sigma_8(z)$ & Ref. \\ [0.5ex] 
 \hline 
    MGS & 0.15 & $-$ & $-$ & $4.508 \pm 0.135$& $0.527 \pm 0.163$ & \cite{Howlett_2015_fs8} \\ \hline
    
    BOSS Galaxy (low-$z$) & 0.38 &$10.27 \pm 0.151$ & $24.89 \pm	0.582$ ~& $-$ & $0.497 \pm	0.045$ ~& \cite{Alam_2017_SDSS_III} \\ \hline

    BOSS Galaxy (high-$z$) & 0.51 & $13.38 \pm	0.179$ & $22.43	\pm 0.482$ & $-$ & $0.459\pm 0.038$ & \cite{Alam_2017_SDSS_III} \\ \hline

    eBOSS LRG & 0.698 & $17.65 \pm	0.302$ & $19.77 \pm	0.469$ & $-$ & $0.473 \pm 0.044$ & \cite{Bautista_2020_SDSS_IV} \\ \hline

    eBOSS QSO & 1.48 & $30.21 \pm 0.7891$ & $13.23 \pm 0.466$ & $-$ & $0.462 \pm 0.045$ & \cite{J_Hou_2020_SDSS_IV} \\ \hline

    Ly$\alpha$ QSO & 2.33 & $37.5 \pm 1.2$ & $8.99 \pm 0.19$ & $-$ & $-$ & \cite{du_Mas_des_Bourboux_2020_SDSS_IV} \\
 
  \hline \hline
\end{tabular}
\caption{SDSS IV data used in our analysis. The covariance information is taken from \url{https://svn.sdss.org/public/data/}.}
\label{tab:sdss}
\end{table*}

The key motivation to study the dynamics of cosmic evolution in the abstract $(x,p,f_8)$ space comes from observations. 
Baryon Acoustic Oscillation (BAO) studies measure the angular scale $\theta_s = s[ (1+z) D_A]^{-1}$ and a redshift interval $\Delta z_s = sH(z)/c$ from the oscillatory signature in the transverse and radial clustering of tracers, respectively. Using a standard ruler $s= r_d$ (the sound horizon at the drag epoch $z_d$), this allows a simultaneous measurement of $D_A(z)$ and $H(z)$. 
Further, peculiar velocity, which introduces additional redshift, leads to an anisotropy in the clustering of tracers when seen in redshift space. If the peculiar velocities are sourced by dark matter overdensities, the redshift space distortion (RSD) imprints the growth rate of density fluctuations $f(z)$. 
The linear growth rate of structure $f(z)$ is not measured independently in tracer clustering because of the degeneracy between $f(z)$ and the tracer bias $b$. 

The amplitude $A$ of the redshift space anisotropic power spectrum is related to $b\sigma_8(z)$, and the redshift space distortion parameter is given by $\beta = f/b$. Thus, even if $A$ and $\beta$ are measured, one would still additionally need either $\sigma_8(z)$ or $b$ to measure $f$. Thus,  measurement of redshift space clustering allows us to measure the quantity $\beta \times \sigma_8(z) b = f_8(z)$. The growth rate $f_8(z)$ depends on $\sigma_8(z)$. If the matter power spectrum is normalized using $\sigma_8$, then the normalization is sensitive to the smoothing scale $R = 8h^{-1} \text{~Mpc}$. It is important to note that the smoothing scale itself depends on $h=H_0/100$. The value of $H_0$ estimated using different models or cosmological probes may differ, providing different physical scales ($8h^{-1} \text{~Mpc}$) and thereby affecting the power spectrum normalization. The matter power spectrum is known to have degeneracy due to parameters $h$ and $A_{s}$ (the primordial amplitude). This introduces a bias and complicates the comparison of $\sigma_8(z)$ or $f\sigma_{8}(z)$ inferred from different cosmological probes or models. This issue is addressed in \cite{Sanchez_fsigma12_2020}. To evade the problem, the use of another parameter $\sigma_{12}(z)$ instead of the traditional $\sigma_{8}(z)$ is suggested. The newly suggested parameter $\sigma_{12}$ normalizes the power spectrum on the fixed physical scale of 12 Mpc and is thus free from the bias introduced by $h$. Thus, $f\sigma_{12}(z)$ can be used as a more robust parameter in RSD analysis instead of $f\sigma_8(z)$ \cite{Sanchez_fsigma12_2020, Forconi_2025_fsigma12}.

In this work, we have still used $f\sigma_8(z)$ instead of  $f\sigma_{12}(z)$ since we are using RSD data from SDSS, which only provides measurements for $f\sigma_8(z)$.

In this section, we discuss the data we have used in our analysis from the BAO and RSD probes. We deliberately considered data where the covariance information between the three key observables $x$ (related to $D_A$),  $p$ (related to $H(z)$), and $f_8(z)$ is available. We have thus tried to use the full covariance information between the observables.

We have used the data from the Sloan Digital Sky Survey IV (SDSS-IV) extended Baryon Oscillation Spectroscopic Survey (eBOSS) \cite{Alam_2021_SDSS}. This data set provides measurements of observables $\widetilde{D}_M(z)$, $\widetilde{D}_H(z)$, $\widetilde{D}_V(z)$ and $f\sigma_{8}(z)$ along with their correlations in the redshift range $0.15 \le z \le 2.33$ \cite{J_Hou_2020_SDSS_IV, du_Mas_des_Bourboux_2020_SDSS_IV} where
\begin{eqnarray}
\widetilde{D}_{M}(z) = \frac{c}{r_d} \int_0^z \frac{dz'}{H(z')}, ~~~ \widetilde{D}_{H}(z) = \frac{c}{H(z) r_d} ~~~ \text{and} \nonumber \\ 
\widetilde{D}_{V}(z) = \left[\frac{(1+z)^2D_A^2(z)cz}{r_d^3H(z)}\right]^{1/3}. ~~~~~~~~~~
\end{eqnarray}
The measured quantities are converted to our dynamical variables using 
\begin{eqnarray}
x(z) &=& \frac{H_0}{c}\frac{1}{1+z}\widetilde{D}_M(z)r_d \nonumber
\\
p(z) &=& \frac{H_0}{c}\frac{r_d}{1+z}\left[\widetilde{D}_H(z) - \frac{\widetilde{D}_M(z)}{1+z}\right].
\end{eqnarray}
The covariance matrices are transformed using the Jacobian of the transformation.
In this work, we  have adopted 
\be r_d = \int _{z_d} ^\infty \frac {c_s(z)  dz}{ H(z)}  = 147.21 \pm 0.23 \text{~Mpc} \ee from CMBR observations \cite{PLANCK18_COSMO_params_Aghanim_2020}. The data used in this work and their sources are summarized in Table \ref{tab:sdss}. We use the covariance matrix between the given observables of the background cosmology and structure formation from a single data source to avoid any tension owing to differences between systematics between two or more different probes. We have also restricted ourselves to the SDSS IV data only, excluding supernovae and cosmic chronometer data, to avoid biasing our analysis towards probes of background cosmology. Any inference we make in this work thus pertains solely to the full BAO + RSD measurements by SDSS.

\section{Method Validation using mock data}

\begin{figure*}[ht]
\centering
\includegraphics[height=10cm, width=18cm]{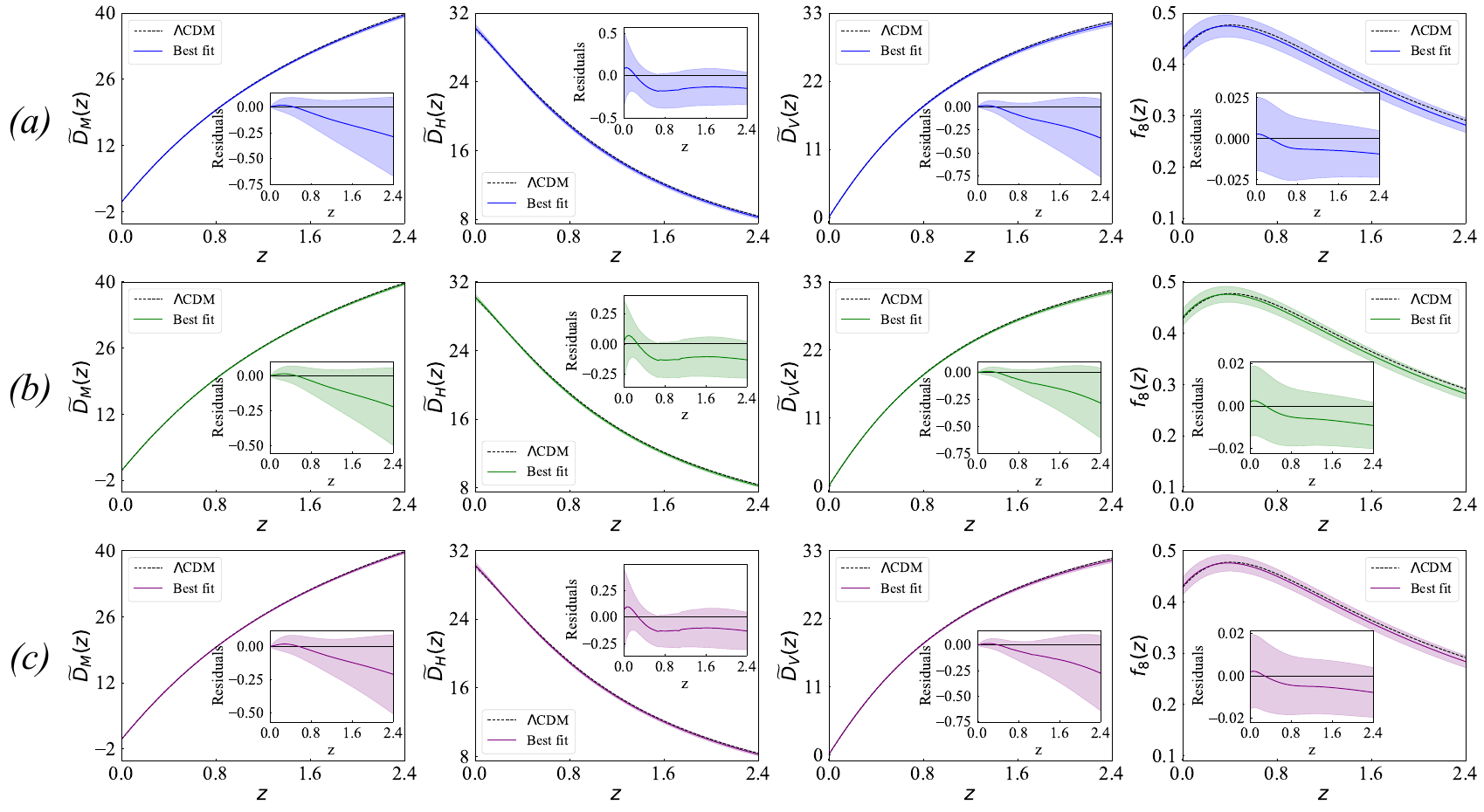}
\captionsetup{justification=justified, singlelinecheck=false}
\caption{Reconstruction of BAO+RSD observables using the semi-cosmographic method on mock data sets. In the insets, we show the residual after subtracting the mean $\Lambda$CDM from the reconstructed functions. The upper panel (a) corresponds to the reconstruction using mock data, where the data is drawn from a multivariate distribution with a mean corresponding to the $\Lambda$CDM model, and the covariance matrix is taken from SDSS IV.  In the second panel (b), the covariance matrix is reduced to 50\% of the original, to see the reduction of reconstruction error. In the lower panel (c), we have introduced data at two more intermediate redshifts to study the effect of data sparsity and additional data.}
\label{fig:residuals}
\end{figure*}

In order to solve for the dynamical quantities ($x(z)$, $p(z)$, $\sigma_8(z)$, $f_8(z)$),  a dark energy equation of state $w(z)$ is required. In a model agnostic approach, no specific $w(z)$ is assumed {\textit a-priori}. In our semi-cosmographic approach,  we use a Pad\'e expansion for the Luminosity distance $D_L^{\cal P}(z)$ of the form given in Eq. \ref{eq:pade_R} \cite{Saini_2000}. We use the semi-cosmographic equation of state in Eq. \ref{eq:eos}, which is then used to solve the dynamical system $(x(z), p(z), \sigma_8(z), f_8(z))$ (see Eq. \ref{eq:autonomous_bgrnd_evln} and Eq. \ref{eq:autonomous_structure_frmn}). The quantities $\widetilde{D}_{M}(z)$, $\widetilde{D}_{H}(z)$, $\widetilde{D}_{V}(z)$ obtained using $x(z)$ and $p(z)$, and $f_8(z)$ 
depend  on the Pad\'e parameters $(\alpha, \beta, \gamma)$  as well as $(H_0, \Omega_{m0}, \sigma_8)$. These six parameters are fitted with the cosmological data on distance ratios from BAO  and the growth rate data on $f(z)\sigma_8(z)$. The flow chart in Figure \ref{fig:Flowchart} shows this semi-cosmographic scheme.

To test our method before applying to real data, we have generated the SDSS-like mock data on  ($\widetilde{D}_{M}, \widetilde{D}_H, \widetilde{D}_V, f\sigma_8$) by sampling over distributions of these quantities with their mean values generated using the $\Lambda$CDM model with Planck18 parameters and the covariance from the SDSS IV data. With this mock data set, we reconstructed the BAO+RSD observables. 
Figure \ref{fig:residuals} shows the results of the implementation of the semi-cosmographic method on mock data sets.

In Figure \ref{fig:residuals}(a), we find that the reconstructed observables are statistically indistinguishable at $< \sim 1\sigma$ from the corresponding  actual $\Lambda$CDM values (in this case of mock data, the Universe is known and is governed by the $\Lambda$CDM model.) Thus, we may apply our semi-cosmographic analysis to actual data and expect that the reconstructed quantities are indeed representative of the actual Universe.

To investigate if the reconstruction errors decrease with improved data (smaller covariance), we consider reduced errors (say to $50\%$ of the original covariance). We find that the error on the reconstructed quantities decreases, indicating better reconstruction with improved data.

To test the method's robustness against redshift coverage, we introduced two additional redshifts $z = 0.6$ and $z= 1.09$, (interpolated between the SDSS redshifts) and obtained the covariance blocks at these redshifts using interpolation of existing SDSS IV covariance. With this extended data set, we reconstruct the observables. The reduced data sparsity (more data points in the given redshift range) also manifests as improved reconstruction and reduced errors.

\section{Results and Discussion}
As discussed in an earlier section, we have chosen to use the SDSS data for which the covariance information between BAO and RSD quantities is available. We have also put Gaussian prior on the parameter $r_d$ from CMBR observations \cite{PLANCK18_COSMO_params_Aghanim_2020} ($r_d = 147.21 \pm 0.23 \text{~Mpc}$) 
and marginalized over it.

\begin{figure}[ht]
% Image on the right
    \centering
\includegraphics[height=8.5cm, width=8.5cm]{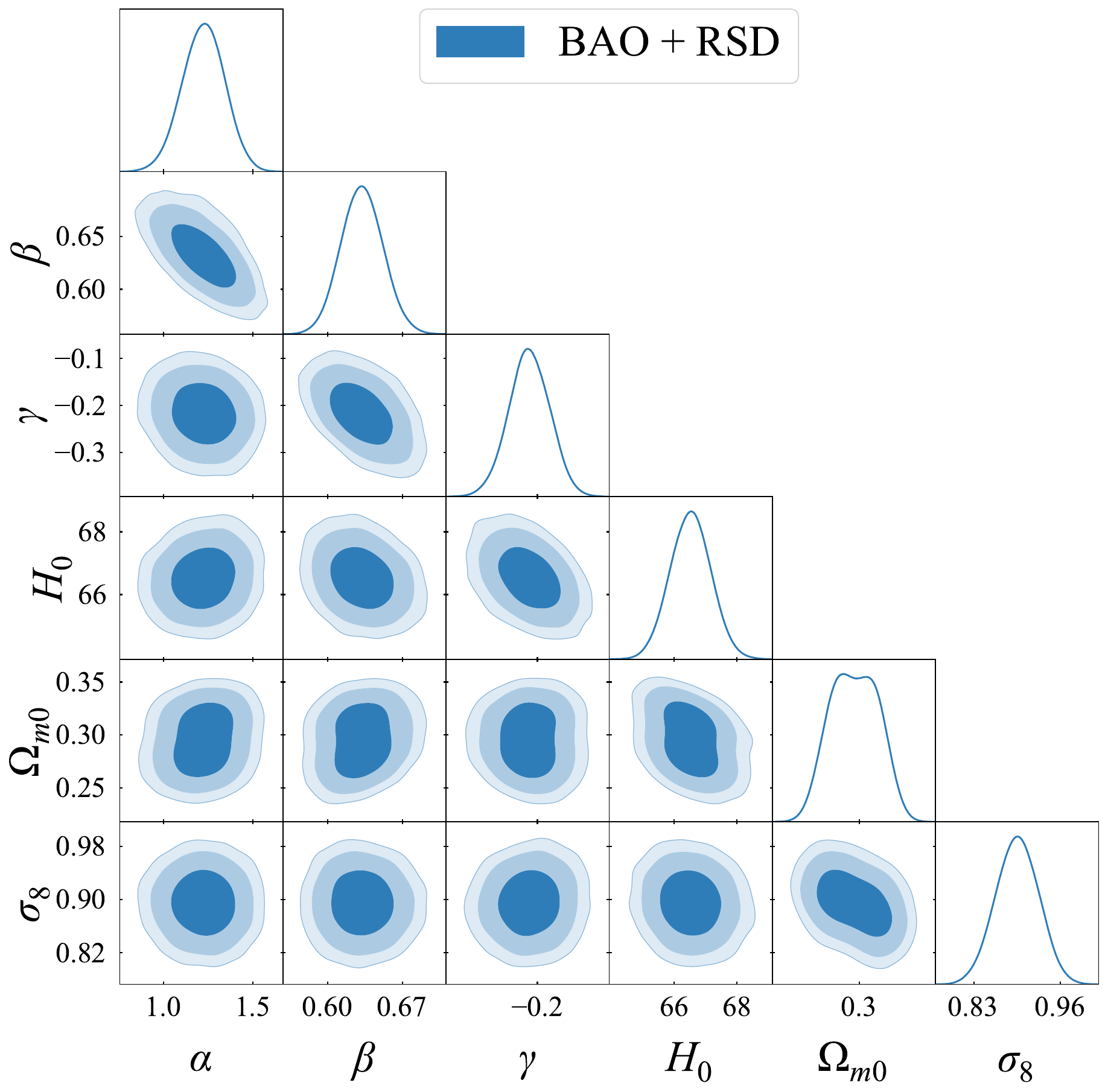}
 \captionsetup{justification=justified, singlelinecheck=false}
 \caption{The posterior distributions of the semi-cosmographic parameters ($H_0, \Omega_{m0}, \alpha, \beta, \gamma, \sigma_8$) after marginalizing over $r_d$. The corresponding 2D correlation plot shows the $68.27\%$, $95.45\%$, and $99.73\%$ confidence contours.}
\label{fig:reconstruct-P}
\end{figure}

To estimate these parameters, we have used the emcee \cite{foreman2013emcee} sampler for our MCMC analysis.
Figure \ref{fig:reconstruct-P}
 shows the posterior distribution of the parameters. 
 The best fit values of the semi-cosmographic parameters $H_0$, $\Omega_{m0}$, $\alpha$, $\beta$, $\gamma$, $\sigma_8$ and $r_d$ along with their $1\sigma$ uncertainty are tabulated in Table (\ref{tab:MCMC-constraints}). The table also shows the best fit values of the $\Lambda$CDM parameters and their corresponding 1-$\sigma$ errors when fitted with the same data. 
 The reduced $\chi_{red}^2 = 1.22$ and $1.41$ for Pad\'e and $\Lambda$CDM, respectively indicate that the fits are good.

We note that the semi-cosmographic Padé approximation using SDSS BAO+RSD data gives $H_0 \simeq 66.5$ km/s/Mpc, while $\Lambda$CDM gives $H_0 \simeq 68.5$ km/s/Mpc. This shift is opposite to the direction needed to alleviate the Hubble tension. On testing the data separately with the CPL \cite{CHEVALLIER_2001} model and an alternative form of Pade expansion given by 
\[ 
    D_L (z) = \frac{c}{H_{0}} \left(\frac{z + a_{2}z^2}{1 + b_{1}z + b_{2}z^2} \right)
\] we find that this trend persists. 

We find that in spite of the different parameterizations and model choices, $H_0$ is only at a $\sim 1-2 \sigma$ tension with the prediction in this work. Thus, we don't see any possibility within these models/model-independent approaches to resolve the Hubble tension.

\renewcommand{\arraystretch}{1.5}
\begin{table*}
\small
\centering
\setlength{\tabcolsep}{2pt}
\begin{tabular}{c c c c c c c c c c c }
%{p{0.15\linewidth}p{0.12\linewidth}p{0.12\linewidth}p{0.12\linewidth}p{0.12\linewidth}p{0.12\linewidth}p{0.12\linewidth} p{0.06\linewidth}}
\hline \hline
Model &$H_0~\rm (Km/s/Mpc)$  & $\Omega_{m0}$  & $\alpha$ & $\beta$ & $\gamma$ & $\sigma_8$ & $\chi^2_{\rm red}$ \\ [0.5ex] 
 \hline 
   
    Pad\'e $D_L^{\cal P} $ & $66.52^{+0.65}_{-0.65}$ & $0.295^{+0.022}_{-0.022}$ & $1.22^{+0.12}_{-0.12}$ & $0.632^{+0.020}_{-0.020}$ & $-0.217^{+0.044}_{-0.044}$ & $0.895^{+0.032}_{-0.032}$ & $1.22$ \\
  $\Lambda$CDM SDSS ~& $68.49^{+0.54}_{-0.54}$ & $0.292^{+0.010}_{-0.010}$ & - & - & - & $0.877^{+0.025}_{-0.025}$ & $1.41$ \\ \hline
  \hline 
\end{tabular}
% \captionsetup{justification=justified, singlelinecheck=false}
\caption{The parameter values obtained from the MCMC analysis marginalized over $r_d$ are tabulated along with the $1\sigma$ uncertainty.}
\label{tab:MCMC-constraints}
\end{table*}

\begin{figure*}[ht]
\includegraphics[height=8.5cm, width=18cm]{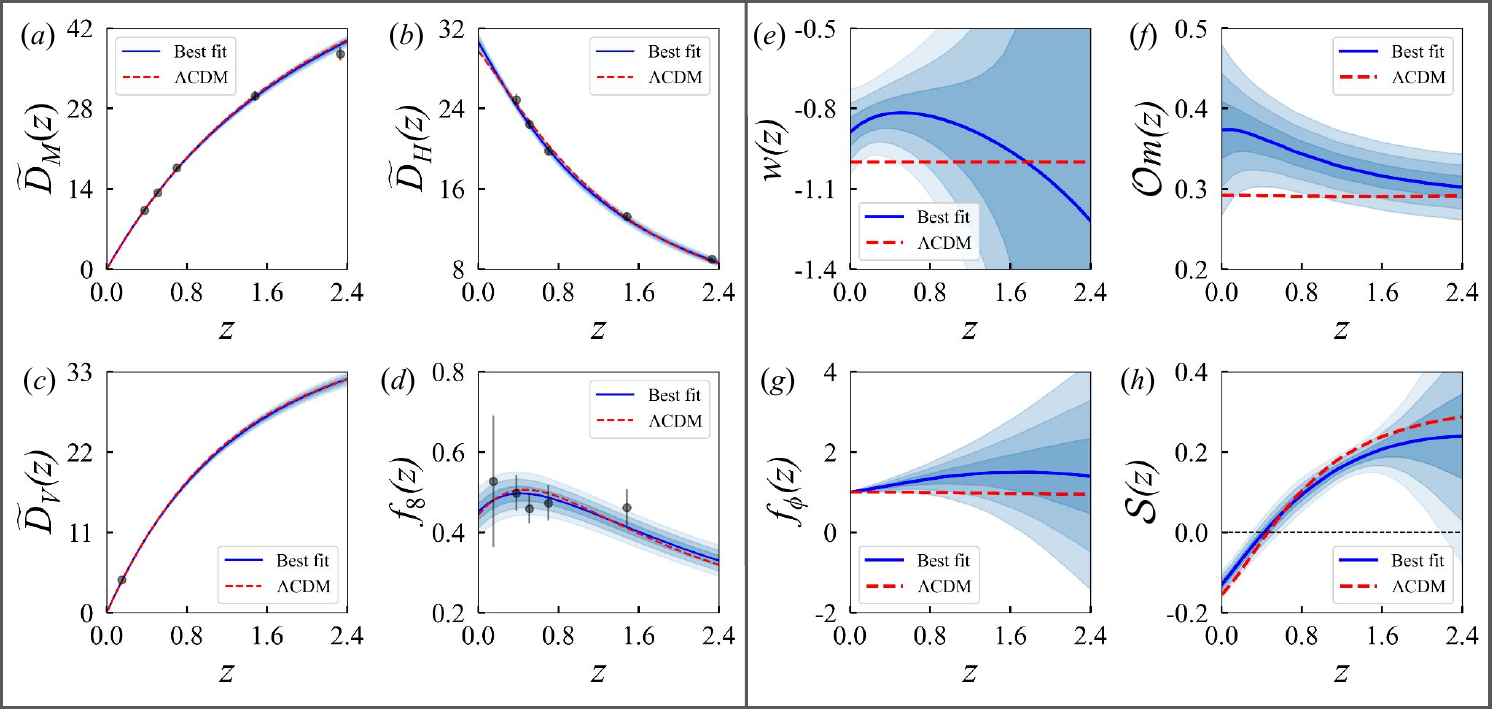}
 \captionsetup{justification=justified, singlelinecheck=false}
\caption{The left part of the panel, figures (\textit{a}), (\textit{b}), (\textit{c}) and (\textit{d}) shows the SDSS data fitted with the semi-cosmographic Pad\'e and $\Lambda$CDM model using MCMC analysis.
The right part of the panel, figures (\textit{e}), (\textit{f}), (\textit{g}), and (\textit{h}) show the reconstruction of some diagnostics of background cosmology for semi-cosmographic Pad\'e and the $\Lambda$CDM model with their $1\sigma$, $2\sigma$, and $3\sigma$ errors.}
\label{fig:diagnostics-recon}
\end{figure*}

\begin{figure*}[ht]
\includegraphics[height=8.5cm, width=18cm]{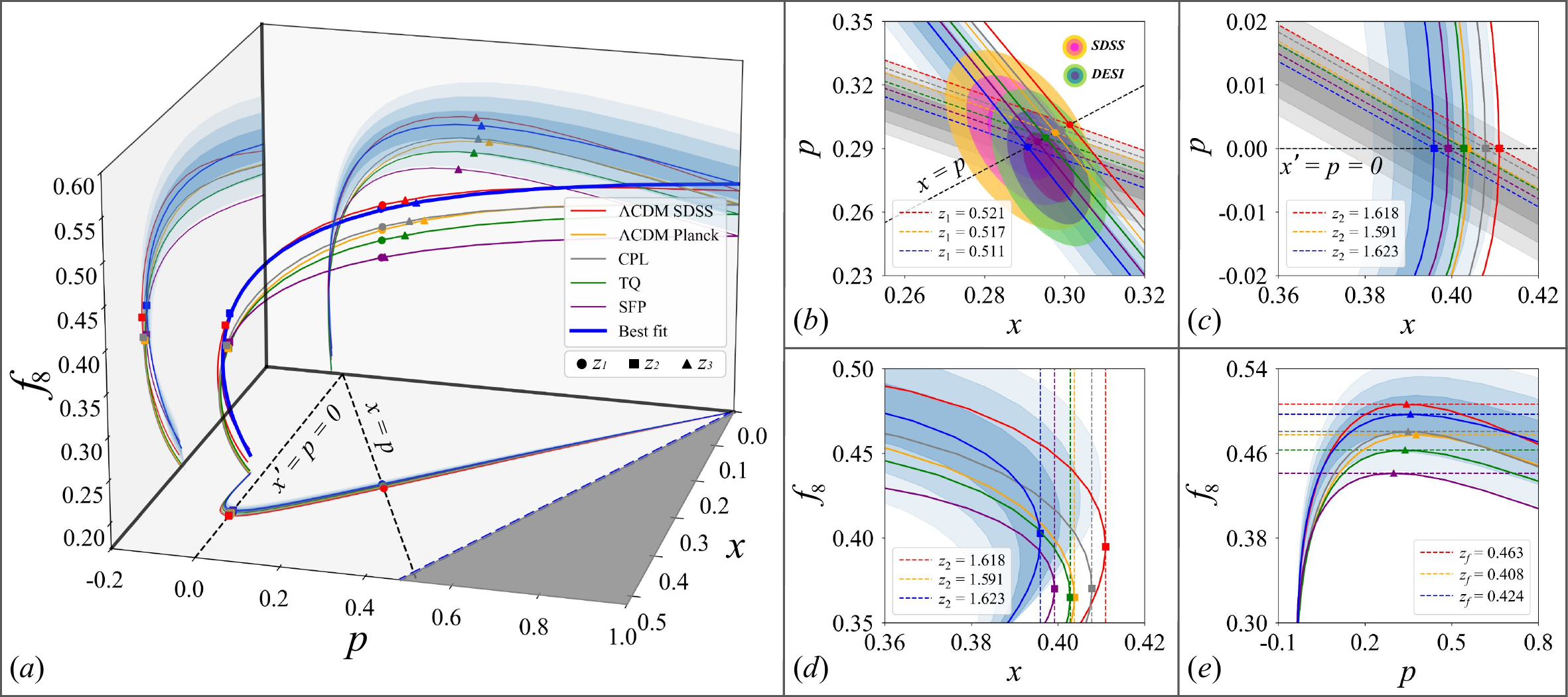}
 \captionsetup{justification=justified, singlelinecheck=false}
\caption{(a) The  reconstructed cosmological evolution in 3-D 
$(x,p,f_8)$ space along with their 
projections on the three places. The $1\sigma$, $2\sigma$, and $3\sigma$ reconstruction errors are shown. 
(b) This shows the part of the $(x, p)$ space where the $x = p$ line intersects the projected phase trajectories. 
We find the reconstructed  unique redshift $z_1$ at which $x(z_1) = p(z_1)$. (c) Shows the projected phase
trajectories in the region of the $(x, p)$ plane where trajectories intersect with the line $p = 0$. The intersection points correspond to the unique redshift $z_2$. (d) Shows the $(x, f_8)$ plane with the marked points corresponding to $z_2$.  (e)
Shows the part of the $(p, f_8)$ space with the marked points on the projected phase trajectories at redshifts $z_f$ where $f_8$ is a maximum.}
\label{fig:3d-recon}
\end{figure*}

Figure \ref{fig:diagnostics-recon} (\textit{a})(\textit{b})(\textit{c}) and (\textit{d}) shows the best-fit reconstruction of the key observable quantities $\widetilde{D}_M(z)$, $\widetilde{D}_H(z)$, $\widetilde{D}_V(z)$ and $f\sigma_{8}(z)$ respectively with the 1$\sigma$, 2$\sigma$ and 3$\sigma$ errors obtained from the MCMC analysis using SDSS data. The actual observational error bars are superposed on the reconstructed results. The best-fit results for the corresponding observables obtained by fitting the  $\Lambda$CDM model with the same SDSS data are also shown for comparison. We find that the $\Lambda$CDM reconstruction is within 1$\sigma$ bounds of the semi-cosmographic reconstruction for all the observables.

Figure \ref{fig:diagnostics-recon} (\textit{e})(\textit{f})(\textit{g}) and (\textit{h}) shows the four dark energy diagnostics $w(z),~\mathcal{O}m (z), ~f_{\phi}(z)$ and $~\mathcal{S}(z)$ defined in Eq. \ref{eq:eos}, \ref{Eq: Om}, \ref{Eq: DE_evolution} and \ref{Eq: s(z)} respectively. Of these, the first three correspond to the background cosmological evolution, and the last one is related to structure formation. 
The corresponding quantities for the $\Lambda$CDM model are also shown. We find that at high redshifts, owing to the large uncertainties, all the diagnostics are indistinguishable from the $\Lambda$CDM model even at 1-$\sigma$. The behaviour of $w(z)$ at low redshifts indicate the $>2.5 \sigma$ 
departure from $\Lambda$CDM value of $-1$, for redshifts $ z < 0.8$.   

The reconstructed $w(z)$ has a divergent behaviour at large redshifts. This pathological behaviour happens whenever the term $E^2(z) / \Omega_{m0}(1+z)^3 \rightarrow 1$. 
This presents a numerical challenge in the semi-cosmographic analysis. One has to impose hard priors to avoid such divergences. However, this would bias the projections. Further, since $w(z)$ is obtained from $D_L^{\cal P}{''}(z)$, each differentiation of a noisy $D_L(z)$ will make the error bars even larger. This results in the error bar on $w(z)$ being very
large, especially at large redshifts, where, owing to a lack of data, the noise on $D_L(z)$ is also large. 

The $\mathcal{O}m(z)$ diagnostic shows $>2.5 \sigma$ departure from the $\Lambda$CDM prediction at low redshifts. At $z<0.7$ the best-fit  $\mathcal{O} m(z)$ satisfies the condition $\mathcal{O}m (z)> \Omega_{m0}$ and thereby, it weakly ($\sim 2 .5 \sigma$) suggests a quintessence dark energy \cite{Sahni_2008_Om}. Like $w(z)$, the quantity $f_\phi(z)$ has large uncertainties at high redshifts, and the current data at high redshift are insufficient to constrain these quantities effectively.

The diagnostic $\mathcal{S}(z)$ (see Eq. \ref{Eq: s(z)}) proposed in this work is related to the dark energy equation of state and the growth rate of density fluctuations. Both the terms appearing in the expression for $\mathcal{S}(z)$ are ratios which are of order unity at large redshifts, and for any $w<0$, $\mathcal{S}(z)$ remains finite over the entire cosmic history $ 0 \leq z \leq \infty$.
This gives the diagnostic $\mathcal{S} (z)$ some advantage over $\mathcal{O}m(z)$ and $w(z)$, since the latter two both have the possibility of divergences. 

Figure \ref{fig:diagnostics-recon}(\textit{h}) shows the reconstructed $\mathcal{S}(z)$ for both the $\Lambda$CDM and semi-cosmographic Pad\'e.
The models are seen to have a $\sim 2\sigma$ tension at low redshifts but are indistinguishable at high redshifts. It is seen that the error in  $\mathcal{S}(z)$ remains small up to high redshifts, where uncertainties on $w(z)$ or $f_\phi(z)$ tend to blow up. The errors on $\mathcal{S}(z)$ at even larger redshifts occur not due to some intrinsic pathology like in $w(z)$ but due to the simple lack of high redshift data. This makes $\mathcal{S}(z)$ a potentially powerful dark energy diagnostic at high redshifts. 

The zero of the $\mathcal{S}(z)$ function corresponds to the redshift $z_f$ where the curve intersects the redshift axis. At this redshift $f'_8(z_f) =0$ and $f_8(z)$ reaches its maximum. At $z=z_f$,  the logarithmic growth rate $f$ is expressible in terms of the background quantities $E(z), \Omega_{m0} $, and dark energy property $w(z)$. Thus, at this redshift $f_8(z_f)$ and its error are completely determined by the measurements of the background observables. This makes $z_f$ a key redshift that should be probed for cosmological investigations. We find that for the semi-cosmographic analysis, $z_f = 0.424$. For the $\Lambda$CDM model fit, we have $z_f = 0.463$. 

We note that in the linear regime the redshift dependency of $\sigma_{12}(z)$ and $\sigma_8(z)$ arises from $f(z) = d\ln D_+/ d\ln a$. Hence, barring an overall amplitude (determined by the normalization), the peak redshift $z_f$ should not be affected if an alternative normalization is used. If the spatial part of the fluctuation is inseparable from the temporal part, there may be a very weak dependency of the peak redshift on the normalization. However, uncertainties in the given data are presently far too large to account for such a weak modification.

Figure \ref{fig:3d-recon}(a) shows the three-dimensional reconstruction of the phase trajectory in the $(x,p,f_8)$ space in the redshift range $0 \leq z \leq 3$. The best-fit portraits are shown for the Pad\'e semi-cosmography and $\Lambda$CDM models. The phase portraits are also shown for some other models (with fiducial parameters taken from the literature). The phase portraits occupy a compact region of the phase space. They start on the $f_8$ axis at $z=0$ and turn around to reach the origin at $z \rightarrow \infty$.
The figure also shows the projections of the phase trajectories on the $(x, p)$, $(x,f)$
and $(p, f)$ planes. The reconstruction errors are shown around the best-fit projected paths for a closer look at possible departures from $\Lambda$CDM.

Figure \ref{fig:3d-recon}(b) shows the zoomed-in part of the $(x,p)$ plane around the region where the $x=p$ diagonal line intersects the phase trajectories. We also show the consistency lines (see Eq. \ref{eq:consistency}). The intersection of these lines (with slope $-(1+z)^{-1}$) with the phase trajectory gives the value $(x,p)$ at that redshift. The $1\sigma$, $2\sigma$, and $3\sigma$ errors on the cosmography reconstructed trajectory are shown. The error on the reconstructed $E(z)$ manifests as errors on the lines of consistency. We find that in this part of the phase space,  the $\Lambda$CDM   trajectory deviates from the best fitted trajectory for the Pad\'e,  at $>3\sigma$. The $x=p$ line crosses the trajectories at a redshift $z_1$. At this redshift, we have
\be 
E(z_1 )  = \frac{1}{x(z_1) ( 2 + z_1) }\ee
At this crucial redshift, we have the expansion rate to be completely determined by  $x(z_1)$, without requiring the derivative $p$. Measurement based on this relationship is therefore important. Thus, $E(z_1)$ is  directly measured from $x(z_1)$. This does not require any other observation or any specific model.
At the point $z_1$, since $x=p$, the variables are perfectly correlated and the error ellipse should collapse to the $45^{\circ}$ line. The best fit trajectory gives $z_1 = 0.511$.  Since we have DESI DR2 and SDSS measurements at this redshift, we show the corresponding error ellipses on the same figure.

\renewcommand{\arraystretch}{1.35}
\begin{table*}
\small
\centering
\setlength{\tabcolsep}{2pt}
\begin{tabular}{l cc @{\hskip 0pt} cc @{\hskip 0pt} c}
\toprule
\hline
Redshift & \multicolumn{2}{c}{\underline{$z_1 =0.511^{+0.002}_{-0.002}$}} & \multicolumn{2}{c}{\underline{$z_2 =1.623^{+0.019}_{-0.018}$}} & \multicolumn{1}{c}{\underline{$z_f =0.424 ^{+0.048}_{-0.045}$}} \\
Observable & $E(z_1)$ & $x(z_1)$ &$ E(z_2)$ & $x(z_2)$ & $f(z_f)$ \\
\midrule
Pad\'e cosmography & $1.370\pm0.016$ & $0.291\pm0.002$~~ & $2.527\pm0.047$& $~0.396 \pm 0.005$ & $0.687\pm0.055$\\
Planck18 $\Lambda$CDM & $1.331 \pm 0.006$ & $~0.299 \pm 0.001$~~~ & $2.524 \pm 0.024$ & $~0.396 \pm 0.004$ & $~~0.736\pm0.006$~~ \\
Tension ($\sigma$)~ & ~~$2.28$~~ & ~~$3.58$~~ & ~~$0.06$~~ & ~~$0$~~ & ~~$0.89$~~ \\ \hline
\bottomrule
\end{tabular}
 \captionsetup{justification=justified, singlelinecheck=false}
\caption{The reconstructed key redshifts with $1\sigma$ errors and the directly measured observables. The Planck 18  estimates are quoted at the same redshifts to indicate the tension. }
\label{tab:zedds}
\end{table*}

Figure \ref{fig:3d-recon}(c) shows the part of the projected plane where the projected trajectories cross the $p=0$ line. 
We refer to redshifts at which this crossing happens as $z_2$.
At this redshift 
\be E(z_2) = {x(z_2) }^{-1} \ee
Since $p=x'=0$  corresponds to the maximum of $x$, the redshift $z_2$ is where the angular diameter reaches a maximum value. 
This is seen clearly in Figure \ref{fig:3d-recon}(d), where in the $(x,f_8)$ plane, we can see that there is a turnaround after $x$ reaches its maximum at $z_2$. At $z_2$, $E(z_2)$ can be directly measured from observed distance $x(z_2)$.

Figure \ref{fig:3d-recon}(e)  shows the projected phase portrait in the $(p, f)$ plane. 
The logarithmic growth rate $f(z) $ grows and saturates to $f(z) \sim 1$ at high redshifts, while $D_+ \sim  \frac{1}{1+z}$ at those redshifts. Thus, the combination $f_8(z) $ exhibits a peak. This behaviour is seen for all models. The peak in $f_8(z)$ occurs at $z=z_f$ where $\mathcal{S}(z_f)$ vanishes. Determining this redshift allows us to measure $f(z_f)$ by assuming $w(z_f)$ and $E(z_f)$  from background cosmology (already constrained from the BAO data). This may be very useful, since $f_8(z)$ is the usual observable quantity in RSD and $f(z)$ is usually inferred using some other data like weak-lensing. However, at $z_f$, one can measure $f(z_f)$ directly by computing background quantities like $E(z)$ and $w(z)$. This may have important observational implications.

Table \ref{tab:zedds} show the three key redshifts $z_1$, $z_2$ and $z_f$. The redshifts $z_1$ and $z_2$ correspond to the points in $x-p$ space where lines $x=p$ and line $x'=p=0$ intersect the projected phase trajectory respectively. The errors in redshift $z_1$ and $z_2$ are obtained by considering the intersection of these lines with the curves corresponding to the reconstructed upper and lower $1\sigma$ uncertainty bounds of the phase trajectory. Similarly, the uncertainty in redshift $z_f$ is obtained by considering the derivative of the reconstructed upper and lower  $1\sigma$ uncertainty bounds of the function $f_8(z)$ in $f_8-p$ space satisfying the condition $\frac{d f_8}{dp}|_{z_f} =0$.

The projected estimate of $E(z_1)$ from our best fit projection is at a $\sim 2.28\sigma$ tension with the Planck projection at the same redshift.
$x(z_1)$ also is at  $\sim 3.58\sigma$ tension with the Planck projection for the same  redshift.
$f(z_f)$ shows  $\sim 0.89\sigma$ tension with Planck projection at the same redshift. This is statistically insignificant. The large errors in the reconstructed $w$ are responsible for this. We also compare our reconstructed $f(z_f)$ with the $f(z_f)$ obtained from the formula in Eq. \ref{eq:lcdmf}, where $\Omega_{m0}$ is adopted from our data analysis. These two results agree at $0.34\sigma$.

We find that at redshift $z=z_2$  the reconstructed observables match with Planck predictions, while such a good match is not seen at $z=z_1$. This might be because at high redshifts, dark energy has little influence on cosmic evolution.  
While the best fit values agree with predictions from other methods and data sets \cite{Purba_mukherjee_2025_Geometric}, they show lesser tension owing to larger reconstruction errors. We are also only using SDSS data as opposed to the other data used in \cite{Purba_mukherjee_2025_Geometric}. 
While our tension with Planck projections is small, we note that such tensions have been reported in projections using Gaussian process reconstruction \cite{Purba_mukherjee_2025_Geometric} where, similar to our work, larger tension is reported at $z=z_1$ and relatively smaller tension at $z=z_2$. Given the completely different reconstruction method and underlying data, this pattern appears to be model independent, and the discrepancy may point towards new physics and is unlikely to be dependent on any reconstruction method. The results are also checked to be stable under the choice of broad priors.
Since the best fit values are quite separated as regards just Planck errors, improved data may sharpen the tensions indicated weakly in this work. 
Our analysis suggests that cosmological observations should be directed towards these key redshifts.

\section{Summary and conclusion}
Motivated by galaxy clustering data, which allows us to measure $D_A(z) $, $H(z)$ (from BAO), and $f\sigma_8$ (from RSD), we have studied cosmological dynamics (both background evolution and growth of structures) in a 3-D $(x,p,f_8)$ space. This approach directs our attention to the idea that comparing cosmological models would necessarily require us to visualize background evolution and structure formation simultaneously. Instead of the traditional practice of seeing each of the functions $x(z)$, $p(z)$, and $f_8(z)$ separately as a function of redshift, we have eliminated the $z$ between them and consider them as a triplet, to be always seen together.
All possible theoretical cosmological models are curves in the accessible region of this phase space which merge at $z = 0$ and $z = \infty$. For all cosmological models, the phase trajectories remain well-behaved and bounded in a finite part of the phase space. 
 Any observational data from tracer clustering at any redshift shall be a point in the 3-D phase space with a region of observational uncertainty around it.
Thus, an observational data point can be immediately compared with pre-imagined theoretical models (which
are existing curves in the phase space). This provides a quick, comprehensive way to detect theory-observation tensions and help rule out theoretical models statistically.

Given the immensity of possible dark energy models, we have adopted a weakly model-agnostic method. 
While we have not taken any specific dark energy model, we have incorporated the dark matter sector through a semi-cosmographic $w(z)$.
Using SDSS data where the covariance data between our relevant triplet members is available, we have reconstructed the 3-D phase path.
We have also reconstructed some diagnostics of
background cosmology and structure formation. The low redshift behavior for these diagnostics indicates a departure from the $\Lambda$CDM model. 
We also find that our proposed diagnostic for structure formation $\mathcal{S}(z)$ suffers less from the pathological divergences as in $w(z)$ and can be used over a greater redshift range as a cosmological probe. 

Finally, we have identified some note-worthy redshifts ($z_1$, $z_2$, $z_f$) such that $x(z_1) = p(z_1)$, $x'(z_2) =0$ and $f_8'(z_f)=0$. Not only do these points have nice geometric interpretations, but they also have important observational ramifications. Future observations may be targeted at these redshifts of significance. 
For example, the DESI DR2 \cite{DESI_Colab_2025_DR2} BAO results have improved statistical precision and are expected to tighten the errors of these redshifts. We find that the DESI DR2 gives us $z_1 = 0.516_{-0.002}^{+0.001}$ and $z_2 = 1.618_{-0.008}^{+0.009}$. There is only a weak tension ($2\sigma$ for $z_1$ and $0.24 \sigma$ for $z_2$) with SDSS IV in the absence of the full BAO+RSD analysis, and these quintessential redshifts seem robust. Interestingly, these redshifts fall in the observational bandwidths of upcoming 21-cm radio intensity mapping experiments by telescopes like MeerKAT \footnote{https://www.sarao.ac.za/science/meerkat/} and SKA Mid \footnote{https://www.skao.int/en}, allowing for the potential alternative BAO measurements using HI distribution. We conclude by noting that these redshifts may be some important temporal slices where answers to cosmological anomalies may be sought.

\section*{Acknowledgments}
AAS acknowledges the funding
from ANRF, Govt of India, under the research grant no.
CRG/2023/003984.

\bibliographystyle{apsrev4-2}
\bibliography{references}
\end{document}